# SPECTRUM PREDICTION AND INTERFERENCE DETECTION FOR SATELLITE COMMUNICATIONS

*Lissy Pellaco[1], Nirankar Singh[2], Joakim Jaldén[1*]*

[1]*Department of Information Science and Engineering, KTH Royal Institute of Technology, Stockholm, Sweden*
[2]*Swedish Space Corporation, Stockholm, Sweden*
*\*pellaco@kth.se, Nirankar.Singh@sscspace.com, jalden@kth.se*

**Keywords**: LONG SHORT-TERM MEMORY, INTERFERENCE DETECTION, SPECTRUM PREDICTION, MACHINE LEARNING, SATELLITE COMMUNICATIONS

## Abstract

Spectrum monitoring and interference detection are crucial for the satellite service performance and the revenue of SatCom operators. Interference is one of the major causes of service degradation and deficient operational efficiency. Moreover, the satellite spectrum is becoming more crowded, as more satellites are being launched for different applications. This increases the risk of interference, which causes anomalies in the received signal, and mandates the adoption of techniques that can enable the automatic and real-time detection of such anomalies as a first step towards interference mitigation and suppression.

In this paper, we present a Machine Learning (ML)-based approach able to guarantee a real-time and automatic detection of both short-term and long-term interference in the spectrum of the received signal at the base station. The proposed approach can localize the interference both in time and in frequency and is universally applicable across a discrete set of different signal spectra. We present experimental results obtained by applying our method to real spectrum data from the Swedish Space Corporation. We also compare our ML-based approach to a model-based approach applied to the same spectrum data and used as a realistic baseline. Experimental results show that our method is a more reliable interference detector.

## 1 Introduction

Spectrum monitoring, and in particular interference detection, is essential for SatCom operators. Signal interference is a major concern in the satellite community. It negatively affects the communication channel, resulting in degraded Quality of Service (QoS), poor operational efficiency, and ultimately revenue loss [1], [2]. In addition, interference is not a rare event: according to a survey conducted by the Satellite Interference Reduction Group (IRG), in 2013, 93 percent of the satellite industry was affected by interference [3]. This issue has not been resolved yet, and the IRG is still actively working to combat and mitigate interference. Moreover, the satellite frequency band is expected to become more and more congested as more countries are launching satellites for different services, such as surveillance, remote sensing, and weather forecast. With the escalating number of users sharing the same frequency band, the risk of interfering signals increases. However, interference is not always unintentional, e.g., due to bandwidth congestion and equipment failure, but could also be caused by malicious signals deliberately transmitted with the purpose of disrupting receiver operations. In the context of satellite navigation, it is very important to detect the presence of jammers that try to spoof the legitimate signals [4], [5].

For all the reasons stated above, spectrum monitoring and interference detection are one of the hot topics in the satellite research field [2], [4]–[6], as they represent the first step towards interference mitigation and suppression. In this paper, we present an approach based on Machine Learning (ML) that guarantees a real-time and automatic detection and localization, in time and frequency, of both short-term and long-term interference in the spectrum of the received signal at the base station. The proposed approach is universally applicable across a discrete set of different signal spectra.

The adopted detection framework follows a standard paradigm [7]: if the received signal spectrum deviates from the expected normal behaviour, then the anomaly flag is raised. This approach is convenient because it overcomes the issue of unforeseen and new anomalies. Given the current signal spectrum, the next spectrum to be received in the absence of anomalies is predicted by making use of historical spectrum data. Then, the prediction is compared, using a proper metric, to the actual received signal. If the comparison metric exceeds a given threshold then an anomaly is detected. In order to predict the normal behaviour of the signal spectrum, an ML-based approach has been chosen. The motivation behind this choice is that machine learning has been widely used in the literature to perform prediction when dealing with time-series data. In particular, among the various machine learning tools, the Long Short-Term Memory (LSTM) has been selected for its proven efficacy in learning both short-term and long-term temporal correlations [8]–[10]. LSTMs have already been adopted to perform prediction of expected normal behaviour with the ultimate goal to detect faults [11]–[13]. However, being able to correctly detect anomalies of different time duration, quantify their length in time and position in frequency is not a trivial task [11], [12], [14]. Nevertheless, our method is able to differentiate and accurately locate in time and in frequency both short-term and long-term interference.

The remainder of the paper is organized as follows: Section 2 describes the ML-based approach, Section 3 presents the experimental results obtained on real spectrum data provided by the Swedish Space Corporation. Section 4 presents a realistic baseline, applied to the same data, and compares it with our proposed approach. Finally, Section 5 concludes the paper.

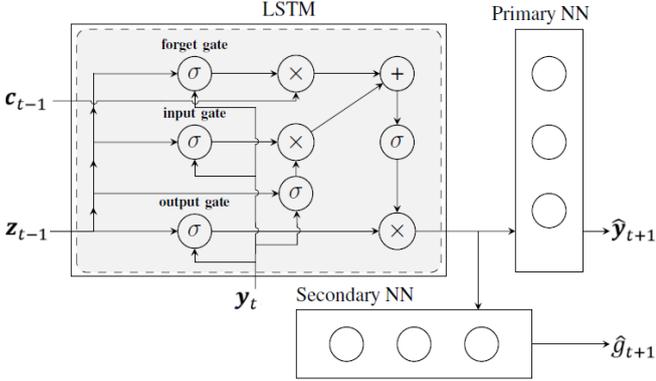

Fig. 1 *Overall system architecture. The prediction of the next spectrum to be received in absence of an anomaly, i.e., $\hat{y}_{t+1}$ is performed given the spectrum at time t, i.e., $y_t$. $c_{t-1}$ is the cell state of the LSTM at the previous time step and contains the relevant information from the past (see (1)-(5) for details). The LSTM is used to perform the prediction and its output is a compact representation of the next spectrum to be received. This compact representation is expanded back to the original dimensionality, i.e., of $\hat{y}_{t+1}$, by the 1-layer fully connected neural network (NN) labelled as "primary". The output of the LSTM is likewise fed into the "secondary" 1-layer fully connected NN that assigns a class label to it, i.e., $\hat{g}_{t+1}$, according to the type of modulation.*

## 2 Proposed Approach

The proposed approach is based on the prediction of the next signal spectrum to be received in absence of anomaly, by using a Long Short-Term Memory (LSTM) trained on historical anomaly-free spectrum data. The prediction is then compared to the actual received signal and, on the basis of a proper designed metric, the anomalies can be detected. In the remainder of this section, the proposed approach will be described in detail and the LSTM structure and working principle will be presented, after defining the adopted notation.

*2.1 Notation and Assumptions*

Vectors are indicated by bold letters, scalars are indicated by regular letters, and matrices are indicated by bold uppercase letters. The power spectral density, in dB scale, at time $t$ is indicated by $y_t$, where $y_t \in \mathbb{R}^d$ is a column vector and $d$=1024, while a single element of $y_t$ is indicated by $y_t[n]$, where $n$ is an integer index in [0,1023]. In the remainder of the paper, for the sake of readability, the power spectral density is referred to as "the spectrum". The various spectra $y_t$ are assumed to be realizations of a correlated and piecewise stationary stochastic process. This assumption, which is proved to be realistic by the experimental results hereinafter, justifies the proposed approach.

*2.2 Method*

Taking into account the temporal correlation among consecutive spectrum realizations in time, like between $y_t$ and $y_{t+1}$, the spectrum at time $t+1$ can be predicted using spectra at time $t, t-1, ..., t-T$. In order to perform the prediction, a Long Short-Term Memory (LSTM) is used. If the training of the LSTM encompasses only anomaly-free data, then the LSTM will learn how to predict anomaly-free spectra. The next step in the direction of anomaly detection is to compare the received spectrum at time $t+1$, i.e., $y_{t+1}$, and its anomaly-free prediction provided by the LSTM, i.e., $\hat{y}_{t+1}$. If an appropriate metric is used for the comparison, then unwanted and unexpected interference/jamming signals can be detected and localized in the frequency spectrum.

Fig. 1 shows the overall system architecture. The output of the LSTM is simultaneously fed into two different 1-layer fully connected neural networks (NN). One of them is labelled as "primary", while the other is labelled as "secondary". The primary network is used to expand the reduced-dimensionality output of the LSTM back to the original input dimension and plays an active role in the anomaly detection task. On the contrary, the role of the "secondary" network is to classify the predicted spectrum into different classes according to the modulation type. This classification does not contribute to the interference detection task; this "secondary" network is added with the aim to gain some insights about the LSTM learning process and outcome. If spectrum classification can be effectively carried out from the reduced-dimensionality output of the LSTM, then we might argue that, during the training process, the LSTM is able to find a compact representation of the signal spectrum containing the relevant features from which the spectrum class can be easily inferred.

*2.3 Long Short-Term Memory*

A Long Short-Term Memory (LSTM) is a type of Recurrent Neural Network (RNN) able to capture and learn short-term and long-term dependencies in sequences of data. The recurrence is represented by the cell state $c_t \in \mathbb{R}^h$ of the LSTM, which is a function of the cell state itself at the previous time step, i.e., $c_{t-1}$, of the LSTM output at the previous time step, i.e., $z_{t-1} \in \mathbb{R}^h$, and of an external input $x_t \in \mathbb{R}^d$. The peculiarity that distinguishes the LSTM from traditional RNNs consists not only in the use of a cell state that stores relevant information from the past, but also in the presence of three so-called gates: the forget gate $f_t$ (see (3)), the input gate $i_t$ (see (4)) and the output gate $o_t$ (see (5)). Such gates are learnable parameters of the LSTM and are respectively used to regulate how much memory from the past should be retained, how to update the cell state with new inputs, and how generate the output from the cell state. This specific structure was designed to mitigate the vanishing/exploding gradient problem [15], which prevents traditional RNNs from learning long-term dependencies, giving the LSTM the ability to learn such long-term correlations [8-10]. The following equations explicit the update rules of the LSTM:

$$c_t = f_t \circ c_{t-1} + i_t \circ \sigma(W_c x_t + U_c z_{t-1} + b_c) \quad (1)$$

$$z_t = o_t \circ \sigma(c_t) \quad (2)$$

$$f_t = \sigma(W_f x_t + U_f z_{t-1} + b_f) \quad (3)$$

$$i_t = \sigma(W_i x_t + U_i z_{t-1} + b_i) \quad (4)$$

$$o_t = \sigma(W_o x_t + U_o z_{t-1} + b_o) \quad (5)$$

where $z_t, c_t, f_t, i_t, o_t, b_{c/f/i/o} \in \mathbb{R}^h$, $W_{c/f/i/o} \in \mathbb{R}^{h \times d}$, $U_{c/f/i/o} \in \mathbb{R}^{h \times h}$, where $\circ$ stands for the Hadamard product, and where $\sigma(\alpha)$ is the element-wise sigmoid function:

$$\sigma(\alpha) = \frac{1}{1 + e^{-\alpha}}$$

The parameters $b_{c/f/i/o}$, $W_{c/f/i/o}$, and $U_{c/f/i/o}$ are learnt by the LSTM during the training process. For further information about the LSTM and its comparison with the RNN, we encourage the reader to refer to [10].

Typically, the cell state $c_t$ stores the relevant information in a compact representation, i.e., $h < d$. Intuitively, the cell state can be thought of as a dimensionality-reduced summary of the significant features of the past input sequences. Therefore, for our purpose, the output of the LSTM at time step $t + 1$, which is a vector in $\mathbb{R}^h$, needs to be expanded back to the original dimensionality $\mathbb{R}^d$, i.e., $\mathbb{R}^{1024}$. This expansion is performed by the "primary" neural network (see Fig. 1). This neural network can be considered as a sort of decoder of the compact representation of the spectrum, implicitly built by the LSTM.

## 3 Experimental Results

*3.1 Dataset*

The data used to run the experiments was provided by the Swedish Space Corporation. It consists of signal spectra belonging to fourteen different passes of satellites. A signal spectrum is represented by a vector of 1024 samples, each representing the power content of the signal at frequency $f$. A pass is the time interval during which the satellite is above the base station so that a communication link can be established. Each pass is characterized by a modulation scheme, either 8PSK or 16QAM, and it contains both spectra of modulated signal and of pure noise. The presence of pure noise spectra is due to the fact that the communication between the satellite and the base station does not necessarily start/end at the exact beginning/end of the pass. The modem parameter "Frame Synchronization" can be used to track the actual start/end of the communication (see Fig. 2). Please refer to Table 1 to have a complete characterization of each pass in terms of number of spectra per pass and type of modulation.

Before feeding the data to the ML tool, element-wise normalization is performed on them:

$$y_{\text{norm}}[n] = \frac{y[n] - a}{b} \quad (6)$$

Given the available dataset, $a$ and $b$ are chosen to be -60 (dB) and 60 (dB), respectively, in order to guarantee that the data fed to the LSTM belongs to the range [-1, 1]. For the sake of readability the subscript "norm" will be dropped, but the data is assumed to be normalized.

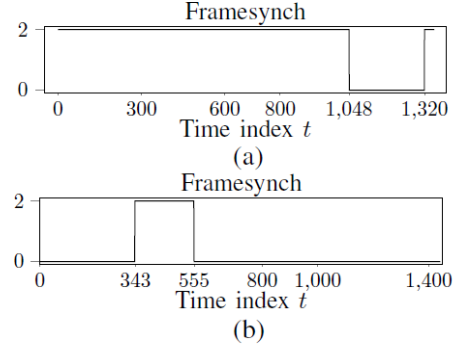

Fig. 2 *Modem parameter "Frame Synchronization" for passes A5 (a) and B5 (b). The value "2" indicates signal transmission between the satellite and the base station, while "0" indicates no transmission.*

Table 1 Dataset provided by the Swedish Space Corporation. It consists of fourteen satellite passes. Each pass is characterized by a different number of spectra and a different type of modulation.

| Pass ID | Number of Spectra | Modulation Type | Pass ID | Number of Spectra | Modulation Type |
|---|---|---|---|---|---|
| A1 | 1241 | 8 PSK | B1 | 1414 | 16 QAM |
| A2 | 1272 | 8 PSK | B2 | 1508 | 16 QAM |
| A3 | 1252 | 8 PSK | B3 | 1416 | 16 QAM |
| A4 | 1252 | 8 PSK | B4 | 1396 | 16 QAM |
| A5 | 1356 | 8 PSK | B5 | 1443 | 16 QAM |
| A6 | 1257 | 8 PSK | B6 | 1351 | 16 QAM |
| A7 | 1197 | 8 PSK | B7 | 1396 | 16 QAM |

*3.2 Architecture and Training*

The overall system architecture is presented in Fig. 1. The output of the LSTM is simultaneously fed into two different 1-layer fully connected neural networks, labelled as "primary" and "secondary". The parameters of the "primary" network are optimized in conjunction with the parameters of the LSTM with the purpose to perform an accurate spectrum prediction. On the contrary, the parameters of the "secondary" neural network are optimized after the LSTM training has ended. The aim of this network is to assign a class label to the output of the LSTM. Four classes are defined according to the spectrum shape: 8PSK-modulation, 8PSK-noise, 16QAM-modulation, and 16QAM-noise. The noise is assigned two different classes because its spectrum changes according to the type of modulation of the pass, as appears in Fig. 3.

The training set is composed of four 8PSK-modulated passes (A1, A2, A3, A4) and four 16QAM-modulated passes (B1, B2, B3, B4). As a result the distribution of the four classes in the training set is the following: 35 percent of the training samples belongs to 8PSK-modulation class, 11.7 percent belongs to 8PSK-noise class, 25.7 percent belongs to 16QAM-modulation class, and 27.6 percent belongs to 16QAM-noise class. The unbalance among the different class types is due to the nature of the provided dataset. As previously mentioned, $d=1024$ and the size of the LSTM cell state $c_t$ and output $z_t$ is chosen to be $h = 20$, i.e., $\mathbb{R}^h = \mathbb{R}^{20}$. The loss function used to train the LSTM and the "primary" neural network is the absolute mean error:

$$l_{\text{abs}} = \frac{1}{I}\sum_{i=1}^{I} |\hat{y}_i - y_i| \qquad (7)$$

where $y_i$ indicates the true $i$-th spectrum, where $\hat{y}_i$ indicates the LSTM prediction of the $i$-th spectrum, and where $I$ indicates the total number of spectra in the training set. The more traditional mean squared error loss, i.e.,

$$l_{\text{mse}} = \frac{1}{I}\sum_{i=1}^{I} (\hat{y}_i - y_i)^2$$

has been tested and compared with $l_{\text{abs}}$ in a preliminary test session. The two loss functions lead to very similar results. However, a significantly smaller training time is required in the case of $l_{\text{abs}}$. Therefore (7) was selected to carry out the experimental analysis. The optimizer adopted is gradient descent, with learning rate of 0.02 and 6000 training epochs. The loss function used to train the "secondary" neural network is the typical cross-entropy loss function in case of multiclass classification:

$$l_{\text{class}} = -\sum_{i=1}^{I}\sum_{r=1}^{R} z_{r,i}\ln(p_r(y_i))$$

where $I$ is the total number of spectra in the training set, where $R$ is the total number of classes, where $z_{r,i} \in \{0,1\}$ has a 1-of-$R$ encoding scheme indicating the true class of the $i$-th spectrum, and where $p_r(y_i)$ indicates the output of the network which is interpreted as the probability that $y_i$ belongs to class $r$. The optimizer adopted is gradient descent, with learning rate of 0.02 and 3000 training epochs. The ML architecture is implemented using Tensorflow [16] and, in particular, its built-in LSTM API.
In order to evaluate the ability to classify the spectra in the test set, the probability of error is computed:

$$P_{\text{error}} = \frac{1}{\tilde{I}}\sum_{i=1}^{\tilde{I}} q(y_i)$$

where $\tilde{I}$ is the number of spectra in the test set after excluding the spectra at the transition instants, i.e., when "Frame Synchronization" goes from 0 to 2 and vice versa, and where

$$q(y_i) = \begin{cases} 1, & g_i \neq \hat{g}_i \\ 0, & \text{otherwise} \end{cases}$$

where $g_i$ is the true class label of the $i$-th spectrum and $\hat{g}_i$ is the predicted class label.

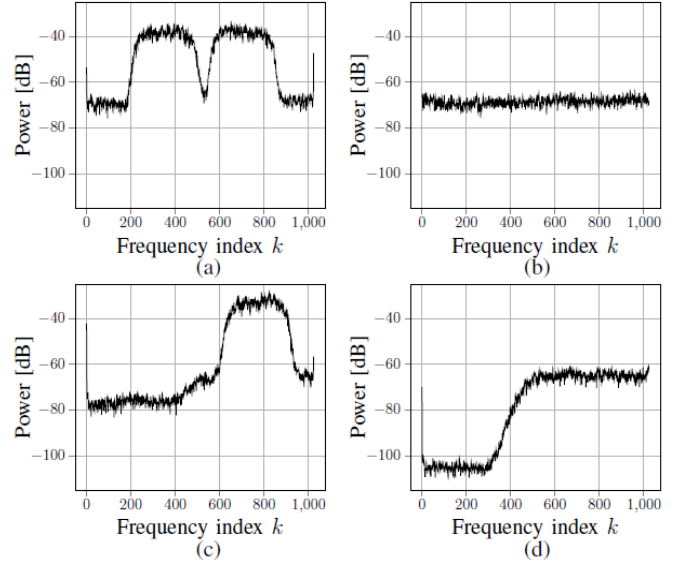

Fig. 3 *Examples of spectra from an 8PSK-modulated pass, in case of signal (a) and noise (b); examples of 16QAM-modulated pass, in case of signal (c) and noise (d).*

In order to test the ability of the proposed method to detect interfering/jamming signals, artificial interferences are manually added to some of the spectra belonging to the test set. Such interferences $y_{\text{int}}$ are column vectors $\in \mathbb{R}^d$ and are chosen to be parabolic in shape and built according to the following equation (in dB):

$$y_{\text{int}}[\varphi] = \begin{cases} -\gamma(\varphi - \beta)^2 + v[\varphi] + \delta, & \varphi \in [0, \Phi] \\ 0, & \text{otherwise} \end{cases} \qquad (8)$$

where $\varphi$ is an integer index in $[0,1023]$, where $\Phi$ indicates the bandwidth of the interference, and where $\gamma$, $\beta$, and $\delta$ determine the amplitude, the center, and the offset of the interference, respectively. $v$ represents Gaussian random noise in the interference and it is drawn from a standard Gaussian distribution, i.e., $v \sim \mathcal{G}(0,1)$.
In order to detect the interference that affects the $i$-th spectrum, the used metric is the Maximum Mean Squared Error (MMSE) computed over a sliding window:

$$\text{MMSE} = \max\{\frac{1}{L}\sum_{j=JL}^{(J+1)L}(\hat{y}_{i,\text{denorm}}[j] - y_{i,\text{denorm}}[j])^2\}$$

where $L$ is the length of the sliding window, $J = 0, 1, 2, \ldots S - 1]$, where $S$ indicates the number of sliding windows, and where $y_{i,\text{denorm}}[j]$ is the $j$-th element of the $i$-th spectrum after removing the normalization specified by (6), i.e., $y_{i,\text{denorm}}[j] = y_{i,\text{norm}}[j]b + a$ with $a$ = -60 (dB) and $b$ = 60 (dB). Analogously, $\hat{y}_{i,\text{denorm}}[j]$ indicates the LSTM prediction of the $j$-th element of the $i$-th spectrum after removing the normalization.

## 3.3 Results

The test set used to validate the proposed method consists of three 8PSK-modulated passes (A5, A6, A7) and three 16QAM-modulated passes (B5, B6, B7) resulting in the following distribution among the four classes: 35.6 percent of the test samples belongs to 8PSK-modulation class, 12 percent belongs to 8PSK-noise class, 14.2 percent belongs to 16QAM-modulation class, and 38.2 percent belongs to 16QAM-noise class. The unbalance among the classes is due to the nature of the provided dataset. Due to limited space we report and discuss only the results obtained on passes A5 and B5. However, these results are good representatives of the entire result set, therefore the considerations and conclusions drawn apply to the entire test set.

The span of the LSTM past memory has been set to one after some preliminary experiments which showed that similar results are obtained even with longer memory spans. It entails that enough information is contained in the spectrum at time $t$ to have a satisfactory prediction of the spectrum at time $t + 1$. Such result derives from the nature of the data and allows to save the additional computational time and power required by an LSTM with longer memory span.

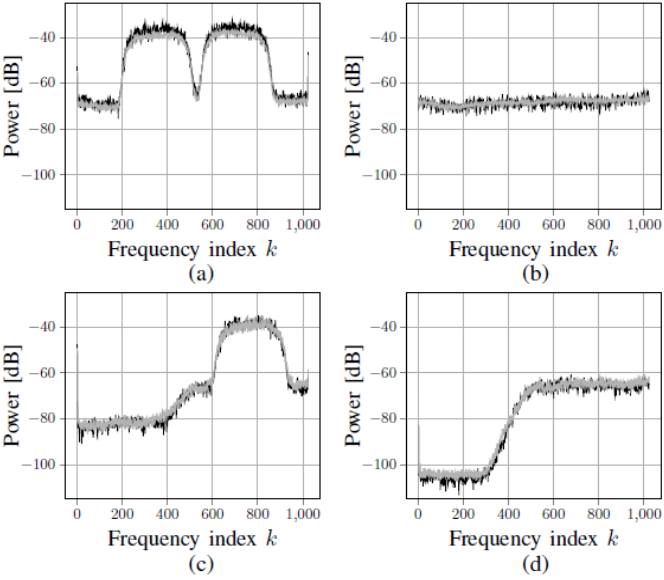

Fig. 4 *LSTM spectrum prediction (in grey colour) versus the ground truth (in black colour), in case of 8PSK modulation (a),(b), and in case of 16QAM modulation (c),(d).*

Fig. 4 shows four examples of LSTM prediction, two in case of 8PSK and two in case of 16QAM, versus their respective ground truth spectra. Note that the LSTM is able to make accurate predictions both in case of noise and modulation. In order to test the ability of the LSTM to detect and locate in time and frequency both short-term and long-term interference a subset of spectra belonging to the test set is altered by adding crafted interference according to (15). The values of parameters $\gamma, \beta$, and $\delta$ in (15) are 40, 0.5, and 20 respectively. The duration of the interference is chosen to be either 10-spectra long (short-term interference) or 100-spectra long (long-term interference). In this scenario the interference alters the spectrum input to the LSTM. Contrary to what one might think, the prediction of the LSTM results to be robust and produces an interference-free spectrum even if the input to the LSTM contains interference itself (Fig. 5). As a consequence, the MMSE is able to highlight the presence of such disturbances, both in time and frequency. Looking at Figs. 6 (a) and (b), it can be immediately noted that two rectangular pulses stand out, together with two spikes. The rectangular pulses identify exactly the two intervals of time in which spectra are altered. The first rectangular pulse from $t = 400$ to $t = 410$ is due to a short-term interference (10-spectra long), while the second pulse from $t = 1100$ to $t = 1200$ is due to a long-term interference (100-spectra long). Conversely, the two spikes, at time $t = 1048$ and $t = 1320$ for pass A5 and at time $t = 343$ and $t = 555$ for pass B5, are due to the transition instants of the "Frame Synchronization" parameter, namely when the communication between the satellite and the base station starts and ends (see Fig. 2 (a) for pass A5 and Fig. 2 (b) for pass B5). In other words, at these two instants, the spectrum shape changes, going from modulation (Fig. 3 (a)/(c)) to noise (Fig. 3 (b)/(d)) and vice versa.

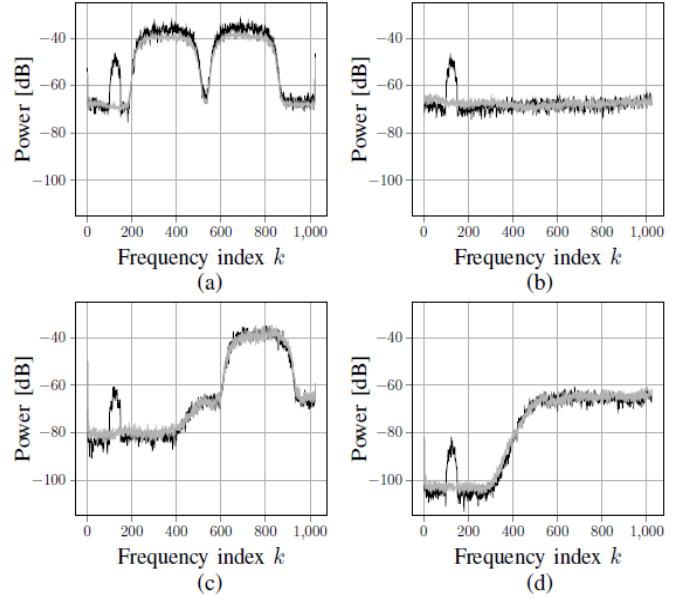

Fig. 5 *LSTM spectrum prediction (in grey colour) versus the ground truth (in black colour), in case of 8PSK modulation (a), (b) and in case of 16QAM modulation (c), (d). The ground truth is altered by an interference (from $k = 100$ to $k = 150$). Note that the prediction performed by the LSTM is not significantly affected by the interference, neither in the case of 8PSK (a), (b) nor in the case of 16QAM (c), (d).*

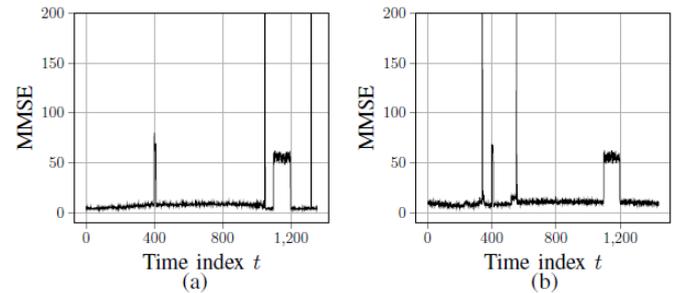

Fig. 6 *MMSE computed over passes A5 (a) and B5 (b). In (a) and (b) the two rectangular pulses at $t = 400$ and $t = 1100$*

*indicate the intervals of time during which the spectrum is affected by interference. Note that the two pulses clearly stick out from the background, meaning that it is extremely easy to detect them. The couple of spikes in (a) at $t = 1048$, $t = 1320$ and in (b) at $t = 343$, $t = 555$ indicate the transition instants from signal to noise and vice versa.*

Therefore, the fact that the MMSE plot contains the two spikes is expected because the LSTM cannot predict when the signal transmission in going to start, nor when it will end. Moreover, the two spikes can be easily filtered out as they differ by almost two orders of magnitudes, therefore it would be difficult to confuse them with actual interference; a simple threshold is sufficient to detect and cancel them out. Likewise, the two rectangular pluses flagging the presence of interference can be easily detected, as they clearly stick out from the background. With respect to spectrum classification performed by the "secondary" neural network, zero percent of probability of error is achieved, after removing the two spectra at the transition instants ($t = 1048$ and $t = 1320$ in pass A5 and $t = 343$ and $t = 555$ in pass B5). Hence, we can argue that the LSTM is able to build a compact representation of the signal spectrum that contains the relevant features from which the spectrum shape, and consequently the spectrum class, can be inferred. From an operational viewpoint, a single spectrum prediction and classification performed by the LSTM takes less than 1 ms on a processor Intel core i7, making it compatible with real-time requirements of satellite systems.

## 4 Comparison with a Model-Based Approach

As previously mentioned, the number of LSTM hidden units is set to one because experimental results show that enough information is contained in the spectrum at time $t$ to obtain a satisfactory prediction of the spectrum at time $t + 1$. Therefore, it is reasonable to examine our proposed approach with respect to a simpler one-time-step approach. This alternative method is not intended to be a state-of-the-art approach. Rather, it is a realistic baseline that we use as a term of comparison.

*4.1 Notation*

The same notation defined in Section 2.1 is used. The spectrum of an 8PSK-modulated signal at time $t$ is indicated by $y_{t,\text{PSK}}$, while the spectrum of pure noise is indicated by $y_{t,\text{PSK}-\text{noise}}$. Analogous notation is used in the case of 16QAM.

*4.2 Method*

The proposed method consists of applying a linear least squares fit [17], after finding a suitable projection of the spectrum in a lower-dimensionality space, spanned by a finite number of basis functions:

$$\boldsymbol{y}_t = \boldsymbol{A}\boldsymbol{\Theta}_t \qquad (17)$$

where $\boldsymbol{A}$ is a matrix whose columns are the basis functions which are column vectors in $\mathbb{R}^{1024}$ and where $\boldsymbol{\Theta}_t$ is a vector whose features are the weights of the linear combination of basis functions that gives the signal spectrum $\boldsymbol{y}_t$. These basis functions need to be carefully engineered in order to be a valid alternative representation of the spectrum. The linear least squares fit is chosen for its convenient closed form solution which makes it widely used in practice, and for the fact that no probabilistic assumption on the data is required. However, the performance of this approach depends largely on the engineered basis functions. In our context, given the four spectrum classes, i.e., 8PSK-modulation, 8PSK-noise, 16QAM-modulation, and 16QAM-noise, it is intuitive that suitable basis functions are given by averaging the spectra belonging to each class:

$$\boldsymbol{f}_{\text{PSK}} = \frac{1}{I_{\text{PSK}}} \sum_{i=1}^{I_{\text{PSK}}} \boldsymbol{y}_{i,\text{PSK}} \qquad (9)$$

$$\boldsymbol{f}_{\text{PSK}-\text{noise}} = \frac{1}{I_{\text{PSK}-\text{noise}}} \sum_{i=1}^{I_{\text{PSK}-\text{noise}}} \boldsymbol{y}_{i,\text{PSK}-\text{noise}} \qquad (10)$$

$$\boldsymbol{f}_{\text{QAM}} = \frac{1}{I_{\text{QAM}}} \sum_{i=1}^{I_{\text{QAM}}} \boldsymbol{y}_{i,\text{QAM}} \qquad (11)$$

$$\boldsymbol{f}_{\text{QAM}-\text{noise}} = \frac{1}{I_{\text{QAM}-\text{noise}}} \sum_{i=1}^{I_{\text{QAM}-\text{noise}}} \boldsymbol{y}_{i,\text{QAM}-\text{noise}} \qquad (12)$$

where $I_{\text{PSK}}$, $I_{\text{PSK}-\text{noise}}$, $I_{\text{QAM}}$, and $I_{\text{QAM}-\text{noise}}$ are the number of spectra belonging to 8PSK-modulation class, 8PSK-noise class, 16QAM-modulation class, and where 16QAM-noise class, respectively. Therefore, $\boldsymbol{A} \in \mathbb{R}^{1024 \times 4}$ and $\boldsymbol{\Theta}_t$ is a column vector in $\mathbb{R}^4$ $\forall t$.

According to this model, any spectrum can be represented by a linear combination of the four basis functions. With this model in mind and the one-time-step dependence, the weights of the linear combination that gives the spectrum at time $t + 1$ can be obtained by applying a linear least-squares fit to the spectrum at time $t$, leading to the following closed form solution:

$$\widehat{\boldsymbol{\Theta}}_{t+1} = (\boldsymbol{A}^T\boldsymbol{A})^{-1}\boldsymbol{A}^T\boldsymbol{y}_t$$

$\widehat{\boldsymbol{y}}_{t+1}$ prediction is now straightforward:

$$\widehat{\boldsymbol{y}}_{t+1} = \boldsymbol{A}\widehat{\boldsymbol{\Theta}}_{t+1}$$

Also, spectrum classification can be performed using this model-based approach. As said before, the features of $\widehat{\boldsymbol{\Theta}}_{t+1} = [\widehat{\boldsymbol{\theta}}_{t+1,\text{PSK}}, \widehat{\boldsymbol{\theta}}_{t+1,\text{PSK}-\text{noise}}, \widehat{\boldsymbol{\theta}}_{t+1,\text{QAM}}, \widehat{\boldsymbol{\theta}}_{t+1,\text{QAM}-\text{noise}}]$ are the weights of the basis functions:

$$\boldsymbol{y}_{t+1} = \widehat{\boldsymbol{\theta}}_{t+1,\text{PSK}} \boldsymbol{f}_{\text{PSK}} + \widehat{\boldsymbol{\theta}}_{t+1,\text{PSK}-\text{noise}} \boldsymbol{f}_{\text{PSK}-\text{noise}} + \widehat{\boldsymbol{\theta}}_{t+1,\text{QAM}} \boldsymbol{f}_{\text{QAM}} + \widehat{\boldsymbol{\theta}}_{t+1,\text{QAM}-\text{noise}} \boldsymbol{f}_{\text{QAM}-\text{noise}}$$

Therefore, each spectrum is assigned to the class represented by the basis function whose weight, after normalization, is the largest.

## 4.3 Experimental Results

Here, we test the prediction accuracy and the interference detection ability of the simple comparison baseline. We adopt the same experimental setup of Section 3. The training set is used to compute the basis functions by applying equations (9)-(12). For the sake of meaningful comparison, we present the results obtained on the same two passes analysed in Section 3, i.e., passes A5 and B5. Also in this case, results on these passes are good representatives of the entire result set. Looking at Fig. 7, it can be seen that this approach is able to perform accurate predictions together with accurate spectrum classification (zero percent of probability of error, after removing the two spectra at the transition instants $t = 1048$, $t = 1320$ in pass A5 and $t = 343$, $t = 555$ in pass B5).

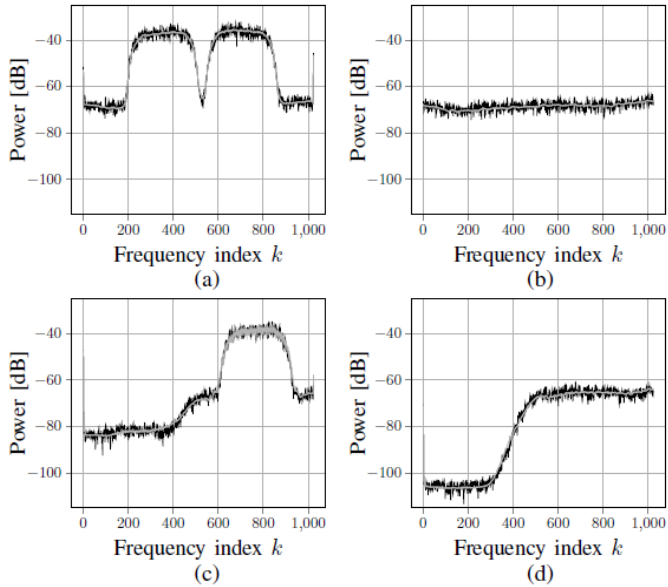

Fig. 7 *Spectrum prediction performed by the comparison baseline (in grey colour) versus the ground truth (in black colour), in case of 8PSK modulation (a), (b) and in case of 16QAM modulation (c), (d).*

However, with respect to interference detection, the comparison baseline does not allow a clear identification of the interference. In Fig. 8, it can be seen that the interference altering the spectrum slightly perturbs the prediction. Even if this minor variation might appear insignificant to the human eye, it results in an MMSE unable to reliably flag interferences: the two rectangular pulses flagging the presence of interference in passes A5 and B5 do not stand out distinctly and are rather complicated to identify (see Fig. 9 (a) and (b)). Conversely, looking at the MMSE obtained with the ML-based approach on the same passes and under the same interference condition, it is immediate to notice both pulses signalling the short-term and the long-term interference starting at $t = 400$ and $t = 1100$, respectively (see Fig. 9 (c) and (d)). It is worth mentioning that in (c) and (d) the background value, which is used as the reference for absence of interference, attains around zero. On the contrary, in (a) and (b) the background value is very different (note that (a) and (b) have a different scale) and it is not even constant in (a), making the use of a threshold for detection problematic, if not impossible. Hence, the presented comparison baseline does not operate as a reliable interference detector.

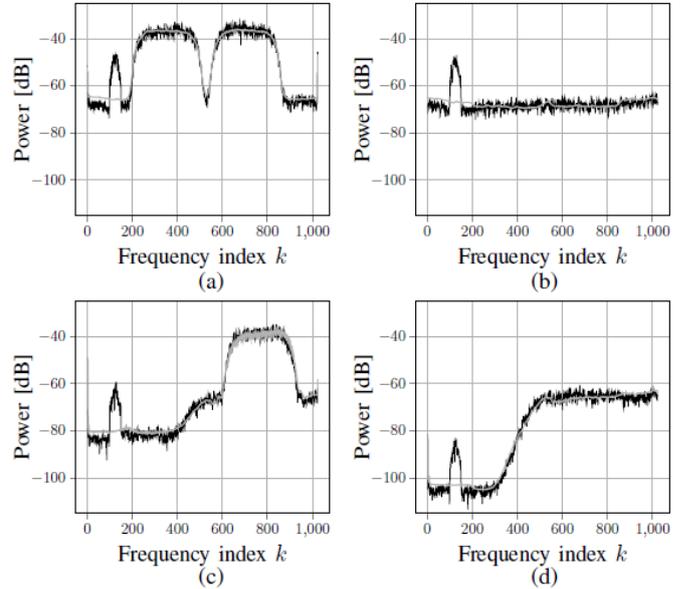

Fig. 8 *Spectrum prediction performed by the comparison baseline (in grey colour) versus the ground truth (in black colour), in case of 8PSK modulation (a), (b) and in case of 16QAM modulation (c), (d). The ground truth is altered by an interference (from $f = 100$ to $f = 150$) that slightly affects the prediction. Even if this minor variation might appear insignificant to the human eye, it results in an MMSE unable to robustly flag interferences.*

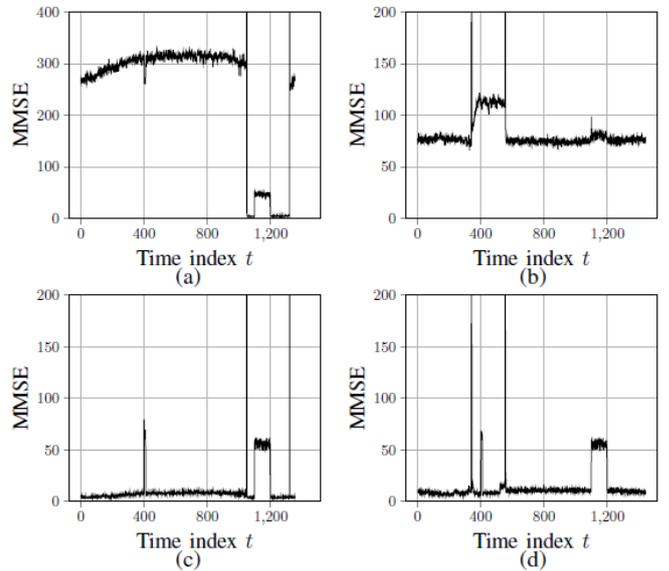

Fig. 9 *MMSE computed over passes A5 (a) and B5 (b) using the predictions performed by the comparison baseline and MMSE computed over the same passes, A5 (c) and B5 (d), using the LSTM predictions. Although in (c) and (d) the two rectangular pulses can be easily detected, in (a) and (b) their identification and localization is extremely difficult, even by human eye.*

*4.4 Comparison*

Both the ML-based approach and the model-based approach prove to be successful at predicting the spectrum at time $t+1$ given spectrum at time $t$. It is an interesting result as the same problem can be effectively addressed in two different fashions. However, each approach has its own benefits and costs. In this context, the model-based approach is easier to implement and deploy as it requires less computational complexity and time. Nevertheless, it might not always be trivial to correctly identify the basis functions which span the subspace of interest to have a complete and rich enough representation of the spectra. Moreover, the training data, in this case, must undergo a heavy pre-processing: it is necessary to divide the data into homogenous categories in order to compute the basis functions, therefore an automatic way to correctly categorize and divide the spectra is mandatory. Conversely, the machine learning-based approach does not require any heavy pre-processing of the data other than the classic normalization procedure. The data can be directly fed into the LSTM as it is, without any categorization or division. Although, it is worth mentioning that the LSTM is a more complex and powerful tool, requiring more computational time and power, thus making implementation and deployment not as straightforward as in the previous case. However, it must not be forgotten that the two approaches are not equivalent: the model-based MMSE plot is unable to robustly identify the presence of the interference leading to a less reliable detection.

## 5 Conclusion

In this paper, we presented a novel ML-based approach that performs real-time and automatic detection of interference in the spectrum of the signal received from the satellite. Furthermore the developed approach is able to accurately locate in time and frequency both short-term and long-term interference and we provided experimental results on real data as evidence of this. Finally, we compared our method with a realistic baseline, against which the ML-based approach proves to be a more reliable interference detector. Future work will be focused on expanding the modulation set considered so far, i.e., 8PSK and 16QAM, to a continuous set which includes not only spectra whose shape fits the expected spectrum shape but also their slightly altered versions.

## 6 Acknowledgements

This project has received funding from the European Research Council (ERC) under the European Union's Horizon 2020 research and innovation programme (grant agreement No 742648), from the Swedish Space Corporation, and from the Swedish National Space Agency under the National Space Engineering Research Programme 3 (NRFP3).